\documentstyle[prb,aps,twocolumn,epsfig,floats]{revtex}
\newcommand{\sS}  { \scriptstyle }

\newcommand{\ket}[1]{ |\!\; {#1} \!\;\rangle} 
\newcommand{\bra}[1]{ {\left\langle\!\: {#1} \!\;\right|} }
\newcommand{\opH} {\hat{H} }
\newcommand{\opD} {\hat{D} }

\newcommand{\vk} {\mathbf{k} }

\begin{document}
\renewcommand{\topfraction}{0.999}
\renewcommand{\bottomfraction}{0.999}
\renewcommand{\textfraction}{0}
\renewcommand{\floatpagefraction}{0.85}
\title{Accessing the dynamics of large many-particle systems\\
using Stochastic Series Expansion}
\author{Ansgar Dorneich}
%\thanks{Also at Physics Department, XYZ University.}
%\email{ardornei@physik.uni-wuerzburg.de}
%\homepage{http://theorie.physik.uni-wuerzburg.de/~ardornei}
\address{Institut f\"ur Theoretische Physik, Universit\"at W\"urzburg,
Am Hubland,  97074 W\"urzburg, Germany}
\author{Matthias Troyer}
\address{Theoretische Physik, ETH Z\"urich, CH-8093 Z\"urich, 
Switzerland
}
%%%%%%%%%%%%%%%%%%%%%%%%%%%%%%%%%%%%%%%%%%%%%%%%%%%%%%%%%%%%%%%%%%
\address{\mbox{}}
\address{\parbox{14cm}{\rm \mbox{ } \mbox{ }
The Stochastic Series Expansion method (SSE) is a Quantum
Monte Carlo (QMC) technique working directly in the imaginary time continuum 
and thus avoiding ``Trotter discretization'' errors. Using a non-local
``operator-loop update''
it allows treating large quantum
mechanical systems of many thousand sites.
In this paper we first give a comprehensive review on SSE and present benchmark
calculations  of SSE's scaling behavior with
system size and inverse temperature, and compare it to the loop algorithm,
whose scaling is known to be one of the best of all QMC methods. 
Finally we introduce a new and efficient algorithm to measure Green's 
functions and thus dynamical properties within SSE.}} 
\address{\mbox{}}
\address{\parbox{14cm}{\rm \mbox{ } \mbox{ }
}}%75.10Nr, 05.30 Jp, 67.40.Yv, 74.60.Ge}}
\maketitle

%%%%%%%%%%%%%%%%%%%%%%%%%%%%%%%%%%%%%%%%%%%%%%%%%%%%%%%%%%%%%%
\section{The SSE technique}
\label{sse_tech}
%%%%%%%%%%%%%%%%%%%%%%%%%%%%%%%%%%%%%%%%%%%%%%%%%%%%%%%%%%%%%%

Since their first formulation in the early eighties\cite{suzu.76,hirs.82}
Quantum Monte Carlo (QMC) methods have become one of the most powerful 
numerical simulation techniques and tools in many-body physics. 
The first QMC algorithms 
were based on a discretization in imaginary time (``Trotter decomposition'')
and used purely local update steps to sample the system's statistically 
relevant states. These methods require a delicate extrapolation to 
zero discretization in order to reduce systematic errors. Furthermore,
the purely local updates often proove incapable to traverse the accessible 
states in an efficient way: autocorrelation times grow rapidly with
increasing system size.

A more recent class of QMC algorithms, the so-called ``loop 
algorithms''\cite{ever.93,hge_more,wiese,Hubbard,nk1,nk2,nk3,LoopReview}, 
use non-local cluster or loop update schemes,
thus reducing autocorrelation times by several orders of magnitude in some 
cases. Unfortunately, it is often highly non-trivial to construct a loop
algorithm for a new Hamiltonian, and some important interactions cannot 
be incorporated into the loop scheme. These interactions have to be added
as {\it a posteriori} acceptance probabilities after the construction of
the loop, which can seriously decrease overall efficiency of the
simulation. Some loop algorithms also suffer from 
``freezing''\cite{ever.93,kohn.97} 
when the probability is high that a certain type of cluster
occupies almost the whole system.

These insufficiencies can be overcome using the ``sto-\-
chastic series expansion''
(SSE) approach together with a loop-type updating scheme 
(see Ref.\onlinecite{sand.99} and earlier works referenced therein).
\begin{itemize}
\item SSE is (almost) as efficient as loop algorithms on large systems.
\item It is a numerically exact method without any discretization error.
\item It is as easy to construct and general in applicability as 
      world-line methods.
\end{itemize}

Following Sandvik\cite{sand.99,sand.92,sand.97} 
we briefly outline the basic ideas of SSE now. 
The central quantity to be sampled in a QMC simulation is the
partition function
\begin{equation}
  Z = \mbox{Tr}(\mbox{e}^{-\beta\opH}),
\end{equation}
where $\opH$ is the system's Hamiltonian and $\beta=1/T$ the inverse 
temperature. 
Standard QMC techniques\cite{vdL.92} split up the exponential 
into a product of many 
``imaginary time slices'' $\mbox{e}^{-\Delta\tau\opH}$ and truncate the
Taylor expansion of this expression after a certain order in $\Delta\tau$,
thereby introducing a discretization error of order $\Delta\tau^n$.
In SSE, however, one chooses a convenient Hilbert base 
$\{\ket{\alpha}\}$ (for example the $S^z$ eigenbase 
$\{\ket{\alpha}\}=\{\ket{S^z_1,S^z_2,...,S^z_N}\}$) and expands $Z$ 
into the power series
\begin{equation}
\label{power_series}
 Z = \sum_\alpha \sum_{n=0}^{\infty} \frac{(-\beta)^n}{n!} \bra{\alpha}
      \opH^n \ket{\alpha}.
\end{equation}
The statistically relevant exponents of this power series are centered around
\begin{equation}
\label{mean_n}
  \langle n \rangle \propto N_s \beta,
\end{equation}
where $N_s$ is the number of sites (or orbitals) in the system. 
(This follows from Eq.\ (\ref{E_mea}) and 
from $\langle E \rangle \propto N_s$.)
We can thus
truncate the infinite sum over $n$ at a finite cut-off length 
$L\propto N_s \beta$
without introducing any systematic error for practical computations. The
best value for $L$ can be determined and adjusted during an initial 
thermalization phase of the QMC simulation: beginning with a relatively small
value of $L$ one can start the QMC update process, stop it whenever the cut-off
$L$ is exceeded and continue with $L$ increased by $10...20\%$. 

Now let $\opH$ be composed of a certain number of elementary interactions
involving one or two sites (such as on-site potentials,
nearest neighbor hopping etc.). In order to obtain a uniform notation we 
combine those interactions affecting only one site to new ``bond'' 
interactions. (One can, for example, take two chemical potential terms 
$\mu\cdot \hat{n}\mbox{\small (site1)}$ and 
$\mu\cdot \hat{n}\mbox{\small (site2)}$ and form the bond
term $\frac{1}{C}\mu(\hat{n}\mbox{\small (site1)}
+\hat{n}\mbox{\small (site2)})$ with the
constant $C$ assuring that the sum over all new bond terms equals the sum over
all initial on-site terms.) We can thus assume in the following that $\opH$ is 
a finite sum of ``bond'' terms $\opH_b$ and that the operator strings $\opH^n$
in (\ref{power_series}) can be split into terms of the form
\begin{equation}
  \prod_{i=1}^n \opH_{b_i}^{(a_i)},
\end{equation}
where $b_i$ labels the bond on which the elementary interaction term operates 
and $a_i$ the operator type (e.g.~density--density interaction or hopping). 
By introducing ``empty'' unit operators 
$\opH^{(0)}=\mbox{id}$ one can artificially grow
all operator strings to length $L$ and obtain\cite{sand.97}
\begin{equation}
  Z = \sum_\alpha \sum_{\{S_L\}} \frac{\beta^n (L-n)!}{L!}
   \bra{\alpha} \prod_{i=0}^L (-\opH^{(a_i)}_{b_i}) \ket{\alpha}.
\end{equation} 
Here $\{S_L\}$ denotes the set of all concatenations of $L$ bond operators 
$\opH_{b}^{(a)}$ and $n$ is the number of non-unit operators in $S_L$.

If we want to sample the $(\alpha,S_L)$ according to their relative weights
with a Monte Carlo procedure we have to make sure that the matrix element 
of each
bond operator is zero or negative since in order to fulfill detailed balance
we choose the acceptance probability $p$ of a bond interaction 
to be proportional to its negative matrix element. 
This requires however that all matrix elements be non-positive.
Does a simple redefinition of the zero of energy help?
For the diagonal operators we can 
indeed add the same negative constant $C$ to each of them without  
changing the system's properties, and thus make all matrix elements 
negative or zero. Unfortunately,
for the non-diagonal terms an equally simple remedy does not exist. If one can 
show, however, that such a non-diagonal operator must appear pairwise for the
matrix element to be non-zero, its matrix element can be multiplied by $-1$ 
without
changing the physics of the system. (This corresponds to a gauge 
transformation
on all lattice sites with odd parity.) On non-frustrated lattices this trick
is widely applicable, which considerably increases the set of Hamiltonians
suitable for SSE. 
If there are valid world-line configurations carrying an odd number of 
non-diagonal vertices with positive matrix elemet -- which is typical 
for Hamiltonians
and lattices with frustrations -- only the conventional approach of dealing 
with the sign problem helps\cite{hirs.82,taka.86,hata.94}: 
one simulates a new system with the acceptance probabilty $p'=|\:\!p\:\!|$ and
obtains the estimate of a physical quantity $Q$ in the form
\[
  \langle Q \rangle = \frac{\langle Q\, \mbox{sign}\,p\rangle}
  {\langle\mbox{sign}\,p\rangle}.
\]
Unfortunately, $\langle \mbox{sign}\, p \rangle$  tends to zero
exponentially with increasing system size $N_s$ and inverse temperature 
$\beta$, so that the computation time needed to 
achieve a certain accuracy exponentially increases with $N_s \beta$ and
the practically accessible range of system sizes and temperatures is 
rather limited.

%%%%%%%%%%%%%%%%%%%%%%%%%%%%%%%%%%%%%%%%%%%%%%%%%%%%%%%%%%%%%%
\section{Loop updates}
\label{loop_upd}
%%%%%%%%%%%%%%%%%%%%%%%%%%%%%%%%%%%%%%%%%%%%%%%%%%%%%%%%%%%%%%

\begin{figure}
\begin{center}
\epsfig{file=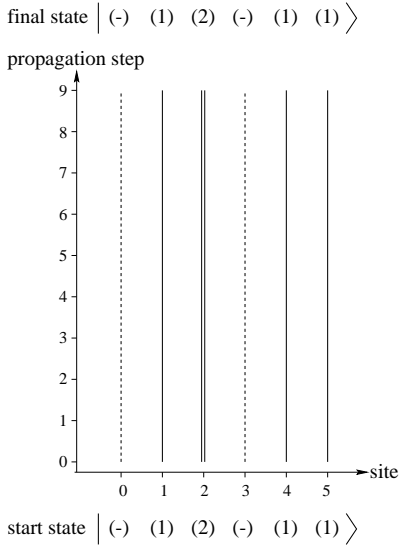,width=5.2cm}
\end{center}
\caption{world-line representation of an arbitrarily chosen start state for
a physical system with three allowed occupations per site: empty (dashed line),
particle 1 (solid line) or particle 2 (double line). 
The initial cut-off length $L$ has been set to $L=9$, and the
initial bond operator string consists only of ``empty'' operators.}
\label{fig1}
\end{figure}

Having outlined the basic idea of SSE we review the non-local updating
updating scheme proposed by Sandvik.\cite{sand.99} 
In the following figures we illustrate the proceeding by means of a 
simple physical model: a system of two types of hard-core bosons 
on a 6-site chain with periodic boundary conditions and Hamiltonian
\begin{eqnarray}
   H&=&-t\sum_{\alpha=1,2}\sum_{i}{\cal P}
   \left[a_{\alpha,i}^{\dag}a_{\alpha,i+1}+H.c.\right] 
   {\cal P} \nonumber \\ && + 
   \sum_{\alpha=1,2}\mu_{\alpha}\sum_{i}n_{\alpha,i}
   \\ &&+ 
   \sum_{\alpha=1,2}\eta_{\alpha}\sum_{i}
   {\cal P}\left[a_{\alpha,i}^{\dag}a_{\alpha,i+1}^{\dag} + H.c.\right]
   {\cal P}\nonumber
\end{eqnarray}
\vspace{-2mm}

with
\vspace{-2mm}
\begin{equation}
{\cal P} = \sum_{i}(1-n_{1,i}n_{2,i})\,.
\end{equation}
The creation operator $a_{\alpha,i}^{\dag}$ creates a hardcore 
boson of type $\alpha=1$ or $2$ on site $i$. The first term ($t$) is 
a nearest neighbor hopping term, the second term ($\mu_{\alpha}$)
the chemical potential and the third term ($\eta_{\alpha}$) pair creation and
annihilation. The projection operator ${\cal P}$ 
implements the hard core constraints between the two types of bosons.
In the world-line representation -- in which
the horizontal axis represents the spatial dimension and the vertical 
axis the propagation
level $l\!=\!1 ... L$ -- we symbolize type-1 bosons by single solid lines, 
type-2 bosons by double lines and empty sites by dotted lines (see Fig.\
\ref{fig1}).
 
Sandvik separates the set of all bond operators
into three classes: empty operators $\opH^{(0)}$, 
diagonal operators $\opH^{(d)}$ and non-diagonal operators $\opH^{(nd)}$. 
The QMC process starts with an arbitrarily chosen initial state $\ket{\alpha}$
and an empty operator string: in Fig.~\ref{fig1}, for example, three sites 
are occupied with type-1 bosons, two sites are empty and on site 2 is occupied
by a type-2 particle. Now two different update steps are performed in 
alternating order: a diagonal update exchanging empty and diagonal 
bond operators 
and an operator loop update transforming and exchanging diagonal and
non-diagonal operators.

\begin{figure}
\begin{center}
\epsfig{file=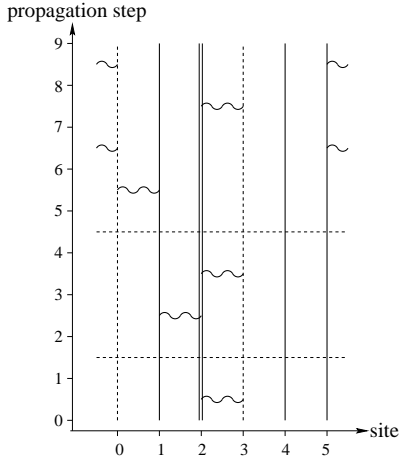,width=5.2cm}
\end{center}
\caption{
In the diagonal update step a certain number of empty bond operators
is replaced by diagonal ones (and vice versa). 
In this example seven of the initial
nine identity operators have been replaced.}
\label{fig2}
\end{figure}

In the diagonal update step the operator string positions $l=1...L$ are 
traversed in ascending order. If the current bond operator is a non-diagonal
one it is left unchanged; if it is an empty or diagonal operator it is
replaced by a diagonal or empty one with a certain probability satisfying
detailed balance (i.e.~an operator with lower energy is more likely to be
maintained or inserted than an operator with higher energy) (Fig.\
\ref{fig2}). 

%=============================================================
%============================================================
%=============================================================
%============================================================

\begin{minipage}{8.3cm}
Following Sandvik\cite{sand.99} we use the notation
\begin{equation}
  \ket{\alpha(l)} = \prod_{i=1}^l \opH_{b_i}^{(a_i)} \ket{\alpha}
\end{equation}
\end{minipage}
for the state obtained by acting on $\ket{\!\alpha\!}$ with the first $l$ bond
operators and $\ket{\!\alpha_b(l)\!}$ for the restriction of 
$\ket{\!\alpha(l)\!}$ 
to the bond $b$. 
Let $M$ be the total number of interacting bonds on the 
lattice. Then the detailed balance conditions for the diagonal update read
\begin{minipage}{7.8cm}
\begin{eqnarray*}
  P(\opH^{(0)}{\sS (l)}\rightarrow \opH^{(d)}_b{\sS (l)}) &&\,= \\
  \min\Big(1 && ,\frac{M\beta\bra{\alpha_b(l)}\opH^{(d)}_b
  \ket{\alpha_b(l)}}{L-n}\Big)\\
  P(\opH^{(d)}_b{\sS (l)}\rightarrow \opH^{(0)}{\sS (l)}) &&\,= \\
  \min\Big(1 && ,\frac{L-n+1}{M\beta\bra{\alpha_b(l)}\opH^{(d)}_b 
  \ket{\alpha_b(l)}}\Big)\,.
\end{eqnarray*}
\end{minipage}
\hfill
\begin{minipage}{5mm}
\begin{equation}
\end{equation}
\end{minipage}
%Resulting numbers larger than 1 are interpreted as 1.

\begin{figure}[h]
\begin{center}
\epsfig{file=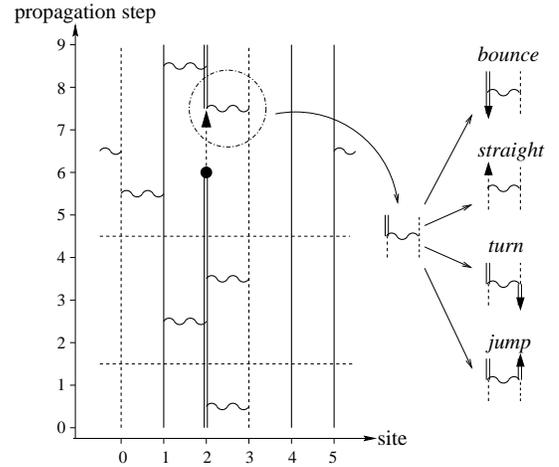,width=7.0cm}
\end{center}
\caption{In the operator loop update step a local change is inserted on a 
world-line and then moved through the world-lines and vertices. At each vertex
a new direction is chosen such that the probability of a path is proportional
to the negative energy of the resulting interaction vertex (detailed balance).}
\label{fig3}
\end{figure}

Non-diagonal bond operators cannot simply be inserted into the world
line configuration as diagonal operators can: their insertion and modification
requires local changes of the world-line occupations. We discussed earlier in 
this paper that
concatenated local changes along a closed path (or loop) through the 
network of world-lines and interaction vertices are much more efficient than
independent purely local changes. Sandvik proposed the following method to
construct such a loop: a certain world-line and a propagation level $l$ on it
is chosen arbitrarily; at the chosen point one disturbs the world-line by a 
local change -- for example the creation or annihilation of a particle. 
Then one chooses a direction (up or down in propagation
direction) and starts moving the disturbation in this direction 
(Fig.~\ref{fig3}). 
The aim is to move this disturbation (we will call it ``loop 
head'' in the following) through the network of world-lines and 
interaction vertices until the
initial discontinuity is reached again and healed up.
%=============================================================
%============================================================

Whenever the loop head reaches an interaction vertex we must decide how to go
on; in the situation shown in Fig.~\ref{fig3} the path ``bounce'' 
is always possible since it results in an unchanged vertex. The path 
``straight'' results in a diagonal vertex, and the path is possible 
if the matrix element of that vertex is nonzero.
The path ``turn'' is only allowed if the Hamiltonian contains 
nearest neighbor hopping terms for particle type 2, while path ``jump'' is 
forbidden unless the Hamiltonian also allows for pair creation of particle 
type 2. The choice among the allowed paths must again satisfy detailed 
balance.

\begin{figure}
\begin{center}
\epsfig{file=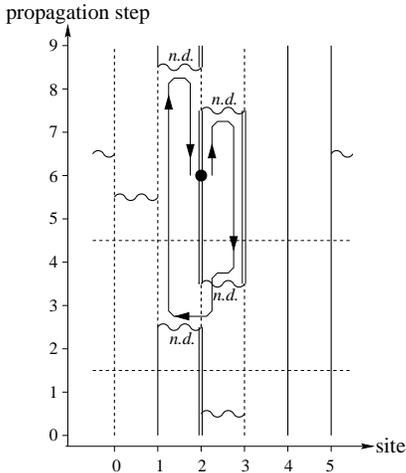,width=5.3cm}
\end{center}
\caption{The loop update closes if the initial insertion point is reached 
again and the inserted world-line discontinuity thus removed.}
\label{fig4}
\end{figure}

In our model -- in which both pair creation and hopping are allowed -- we 
might end up
with the series ``turn'', ``jump'', ``turn'', ``turn'' of path choices, after
which the starting point is regained and the world-line discontinuity healed
up (Fig.~\ref{fig4}). The overall result of this loop is that we have 
replaced 4 diagonal interactions by 4 non-diagonal interactions (marked 
``n.d.'') in Fig.~\ref{fig4}.
%=============================================================
%============================================================

Sandvik's method implicitly assumes that running with a world-line change
into an interaction vertex always requires choosing an outgoing leg and 
a change on it and continuing the loop. But what if the encircled vertex in
Fig.~\ref{fig3} with three empty legs and one leg occupied by particle 2
is also a valid vertex? Then we have to add a fifth possibility to the list
of allowed path choices: the ``stop here''. 
If this last alternative is chosen the loop has reached a dead end. 
In this case our SSE code terminates the loop here, goes back to the starting 
point and moves in the opposite direction until either another dead end is 
encounterd or the starting point is reached again and the initial discontinuity
is healed up.

From Fig.~\ref{fig3} one can see that when choosing a path through the 
current vertex there is always the possibility to undo the current change on
the incoming leg of the vertex and to ``bounce'' backwards on the same
leg. This path choice is normally not very helpful since it means one step
backwards in the construction of the current loop. Fortunately, all 
``bounce'' paths can be suppressed without violating detailed balance 
if on each bond all nonzero matrix elements are equal, or can be made 
equal after a suitable energy shift of the diagonal vertices. As an additional
benefit,
without the ``bounce'' path the algorithm becomes equivalent to the
loop algorithms. For each vertex a path can be chosen
according to detailed balance, after which the loop construction becomes
deterministic. All the Heisenberg models studied 
in section \ref{sec_scal} are examples for this class of ``optimizable''
physical systems.

A further improvement of the update scheme is possible in the limit of 
high temperatures, i.e.~$\beta\rightarrow 0$. Equation 
(\ref{mean_n}) tells us that the average number of (non-empty) vertices is
rather small in this situation, and a large part of all world-lines 
is not connected to any vertex at all. The loop update will not
be very efficient here since it essentially needs a sufficient number of
vertices interconnecting the world-lines. For this reason our SSE code
additionally performs a so-called ``free world-line update'' on each world 
line carrying no vertex at all. In this update the occupation of the entire
world-line is changed to a randomly selected new occupation.  

We have stressed several times that all the local path choices satisfy detailed
balance. What remains to be shown is that the updating mechanism is 
ergodic in the grand canonical ensemble, i.e.~that all bond operator 
strings 
$S_L$ and all states $\ket{\alpha}$ can be reached. 
In order to demonstrate that we remind the reader 
that loops crossing the boundary between first and last 
propagation level $l$ modify the initial state $\ket{\alpha}$ for the next 
update cycle. Therefore, the loops sample not only $S_L$ but also 
$\ket{\alpha}$, and starting from a completely empty system {\em any}  
allowed configuration can be generated by a series of loops traversing one
entire world-line each.

Numerical tests of the loop-update mechanism described above show that for
large system sizes and if there are elementary interactions with
very different energy scales, the loop construction sometimes gets stuck 
and the loop head does not find its way back to the
starting point even after millions of steps. In order to avoid this trapping
loops that exceed a critical length are aborted and the original state
of the vertices is restored. This causes no systematic errors 
for measurements done
between loop updates as detailed balance is not violated. 
The measurements of Green's functions $G(r)$,  however,
which are performed ``on the run'' during loop construction (see Sec. 
\ref{sec_GF}), are biased if large loops are thrown away.
Since the large loops are more likely to reach regions of the systems far away 
from the starting point than short loops, the values of $G(r)$ for large
distances $r$ are systematically under-estimated if a considerable amount
of large loops is aborted. Hence the total number of aborted loops has to
be checked before one can trust in the recorded Green's functions.

%%%%%%%%%%%%%%%%%%%%%%%%%%%%%%%%%%%%%%%%%%%%%%%%%%%%%%%%%%%%%%
\section{Measurements}
\label{sect_meas}
%%%%%%%%%%%%%%%%%%%%%%%%%%%%%%%%%%%%%%%%%%%%%%%%%%%%%%%%%%%%%%

Efficient estimators for many static observables within the SSE mechanism
have been derived by Sandvik.\cite{sand.97a}
\begin{itemize}
\item
All observables $\opH^{(a)}$ appearing as elementary interactions in the 
system's Hamiltonian can be measured very easily by counting 
the corresponding interaction vertices in the bond operator string $S_L$: 
if $S_L$ contains on average $\langle N(a) \rangle$ such vertices one obtains 
\begin{equation}
  \langle \opH^{(a)} \rangle = -\frac{1}{\beta}\langle N(a) \rangle.
\end{equation}
\item
Summing over all elementary terms $\opH^{(a)}$ gives an estimator for the
internal energy $E$:
\begin{equation}
\label{E_mea}
  E = -\frac{1}{\beta} \langle n \rangle,
\end{equation}
where $n$ is the number of non-empty interaction vertices in $S_L$.
(This equation can be derived very easily from 
$\langle E \rangle = \frac{\partial}{\partial\beta}\mbox{ln} Z$.)
\item
For the heat capacity $C_V$ we additionally have to measure the fluctuations
of $n$:
\begin{equation}
  C_V = \langle n^2 \rangle - \langle n \rangle ^2 - \langle n \rangle.
\end{equation}

\item
Equal time correlations of two diagonal operators $\opD_1$ and $\opD_2$ can be
measured via
\begin{equation}
  \langle \opD_1 \opD_2 \rangle = \left\langle \frac{1}{n+1} 
  \sum_{l=0}^{n} d_2[l]\,d_1[l] \right\rangle,
\end{equation}
where $d_i[l]=\langle\alpha(l)|\opD_i|\alpha(l)\rangle$.
\end{itemize}

Are there equally efficient estimators for time-depen\-dent observables?
In SSE the propagation index $l$ describes the evolution of an initial state
when a series of elementary terms of the Hamiltonian is acting on it;
thus $l$ plays a role analogous to imaginary time in a standard path integral.
More detailed calculations\cite{sand.92} show that an imaginary time separation
$\tau$ corresponds to a binomial distribution of propagation distances 
$\Delta l$;
the time-dependent correlation
$\langle \opD_2(\tau)\opD_1(0)\rangle$, for example, is related to the 
correlator
\begin{equation}
 C_{12}(\Delta l) = \frac{1}{n+1} \sum_{l=0}^{n} d_2[l+\Delta l]\,d_1[l]
\end{equation}
via
\begin{equation}
\label{time_dep_corr}
  \langle \opD_2(\tau)\opD_1(0) \rangle \!=\! \left\langle 
  \sum_{\Delta l=0}^n\!\! {n \choose \Delta l}\!\!\left(\frac{\tau}{\beta}\right)^{\!\!\Delta l}\!\!\!
  \left(1\!-\!\frac{\tau}{\beta}\right)^{\!\!n-\Delta l}\!\!\!\!\!
  C_{12}(\Delta l)\! \right\rangle.
\end{equation}
Instead of working in a representation with varying $n$ a fixed string size $L$ can be chosen, as the identity vertices are uniformly distributed and do
not influence the mapping from index to imaginary time.

The corresponding generalized susceptibilities can be calculated straight
forward by integrating $\langle \opD_2(\tau)\opD_1(0)\rangle$ over $\tau$,
\begin{equation}
  \chi_{12}=\int_0^\beta \langle \opD_2(\tau)\opD_1(0) \rangle\, d\tau.
\end{equation}
which gives\cite{sand.92}
\begin{eqnarray}
  \chi_{12} =&& \Bigg\langle {\beta\over n(n+1)}
  \left (\sum\limits_{l=0}^{n-1} d_2[l] \right )
  \left (\sum\limits_{l=0}^{n-1} d_1[l] \right )\nonumber\\
  &&+{\beta\over (n+1)^2} \sum\limits_{l=0}^{n} d_2[l]\,d_1[l] \Bigg\rangle .
\end{eqnarray}

%%%%%%%%%%%%%%%%%%%%%%%%%%%%%%%%%%%%%%%%%%%%%%%%%%%%%%%%%%%%%%
\section{Scaling Behavior}
\label{sec_scal}
%%%%%%%%%%%%%%%%%%%%%%%%%%%%%%%%%%%%%%%%%%%%%%%%%%%%%%%%%%%%%%

One decisive criterion for the performance of a QMC simulation technique
is the behavior of computation time $C$ as a function of system size $N_s$
or inverse temperature $\beta$;
To facilitate a hardware-independent
measurement of $C$ and a comparison to other QMC techniques we define $C$
as the number of elementary update operations needed to transform a 
given state 
$\ket{\!\alpha^{(n)}\!}$ into an new state $\ket{\!\alpha^{(n+1)}\!}$ 
in such a way
that the mean autocorrelation time $\tau$ is equal to 1.
In SSE the number of elementary update operations is the number of diagonal
vertices tested for replacement plus the number of vertices traversed 
during the loop update. 
In the following we compare SSE to the loop-algorithm,
which is known to show an excellent scaling behavior for many benchmark
problems. As test models we choose isotropic antiferromagnetic 
Heisenberg models in one, two and three dimensions 
with up to 4096 sites and $\beta$
up to $64$ in a vanishing or finite external magnetic field.

Following Ref. \onlinecite{li.ma.95} we describe the scaling behavior 
of the two 
algorithms by means of the dynamical exponent $z$ defined from
\begin{equation}
\label{eq_dyn_exp}
  \tau\!\cdot\!C\propto \beta\!\cdot\!l^D l^z\,.
\end{equation}
Here, $\tau\!\cdot\!C$ is the computational effort (i.e.~the number of 
elementary update steps) needed to achieve a mean autocorrelation time of
$\tau\!=\!1$ for the measurements of the studied quantity; $D$ is the 
spatial dimension of the simulated system and $l=\sqrt[D]{N_s}$
its length in each dimension.
From Table 
\ref{tab_scal2Dh0} we see that both simulation techniques 
show an approximately equal performance and an very good scaling behavior:
since the ratio $C\tau/(\beta N_s)$ is approximately constant we obtain
$z\approx 0$ in both cases.

%============================================================
\begin{table}
\caption{2D antiferromagnetic Heisenberg model at vanishing magnetic field
$h=0$: calculation of the uniform magnetic susceptibility 
$\langle\chi\rangle = \frac{\partial \langle M\rangle}{\partial h}\big|_{h=0}$
from QMC simulations with 1000000 (120000 in the case $\beta=L=64$) 
update-measurement cycles (left: SSE, right: loop-algorithm). 
$C\cdot\tau$ is the number of elementary update operations per cycle needed
to achieve a mean autocorrelation time $\tau=1$ for the measurements 
of $\chi$.}
\label{tab_scal2Dh0}
\begin{center}
\begin{tabular}{rclcl}
 & \multicolumn{2}{c}{SSE} & \multicolumn{2}{c}{Loop} \\ 
$\beta\cdot N_s$ & 
$\displaystyle \frac{C\cdot\tau}{\beta\cdot N_s}$ &
\multicolumn{1}{c}{{\normalsize $\chi$}} 
& $\displaystyle \frac{C\cdot\tau}{\beta\cdot N_s}$ & 
\multicolumn{1}{c}{{\normalsize $\chi$}} \\[1.7ex] \hline\\[-2.2ex]
$4\cdot\,\; 4^2$ & 1.00 & $0.040(46\pm 16)$ &  1.00  & $0.040(20\pm 10)$ \\
$8\cdot\,\; 8^2$ & 0.61 & $0.044(83\pm 15)$ & 0.90 & $0.044(92\pm 8)$ \\
$16\cdot 16^2$ & 0.40 & $0.044(72\pm 12)$ & 0.56 & $0.044(68\pm 6)$ \\
$32\cdot 32^2$ & 0.40 & $0.044(19\pm 11)$ & 0.53 & $0.044(24\pm 6)$ \\ 
$64\cdot 64^2$ & 0.36 & $0.044(01\pm 23)$ & 0.42 & $0.044(07\pm 14)$ \\
\end{tabular}
\end{center}
\end{table}
%============================================================

Next we enlarge the square lattice into the third spatial dimension and
examine a bilayer quantum Heisenberg antiferromagnet 
at the quantum critical point separating the spin gap phase 
from the magnetucally ordered one.\cite{bilayer}
Our aim is to measure
scaling behavior and dynamical exponents exactly at this quantum critical 
point. This point is of particular interest since the immediate neighborhood
of a phase transition often leads to the so-called 
``critical slowing down'' of QMC simulations, 
i.e.~exploding autocorrelation times and thus a dramatic decrease
of efficiency of the QMC update process.

The results in Table  
\ref{tab_scalBih0} show that the scaling behavior for
both algorithms is still almost linear in $\beta N_s$. The
scaling for SSE looks slightly superior to the loop algorithm. This
difference can most probably be attributed to the fact that
improved estimators were used in the loop algorithm simulation,
leading to slightly smaller errors but larger autocorrelation times.
There is no sign of critical slowing down in either algorithm.

%============================================================
\begin{table}
\caption{Square bilayer antiferromagnetic Heisenberg model at vanishing 
magnetic field $h=0$ and at the quantum critical point ($J_\perp/J=2.524$):
calculation of the uniform magnetic susceptibility $\langle\chi\rangle$
from QMC simulations with 1000000 (390000 in the case $\beta=L=32$) 
update-measurement cycles.} 
\label{tab_scalBih0}
\begin{center}
\begin{tabular}{rclcl}
 & \multicolumn{2}{c}{SSE} & \multicolumn{2}{c}{Loop} \\
$\beta\cdot N_s$ & \rule[-3.5mm]{0mm}{8.5mm}
$\displaystyle \frac{C\cdot\tau}{\beta\cdot N_s}$ &
\multicolumn{1}{c}{{\normalsize $\chi$}} 
& $\displaystyle \frac{C\cdot\tau}{\beta\cdot N_s}$ & 
\multicolumn{1}{c}{{\normalsize $\chi$}} \\[1.7ex] \hline\\[-2.2ex]
$4\cdot\,\; 2\cdot 4^2$ & 1.00 & $0.0115(6\pm 7)$ &  1.00 & $0.0114(6\pm 5)$\\
$8\cdot\,\; 2\cdot 8^2$ & 0.96& $0.0068(2\pm 6)$ &  1.03 & $0.0069(2\pm 2)$ \\
$16\cdot 2\cdot 16^2$ & 0.68 & $0.0036(8\pm 4)$ & 1.20 & $0.0036(6\pm 2)$ \\
$32\cdot 2\cdot 32^2$ & 0.56 & $0.0018(5\pm 3)$ & 1.20 & $0.0018(3\pm 1)$ \\
\end{tabular}
\end{center}
\end{table}
%============================================================

As we have mentioned in the introduction one of the major 
advantages of SSE is that external potentials 
(and magnetic fields in spin models) can be included
without a loss of performance. To verify this assertion we now examine the
antiferromagnetic Heisenberg model on a chain and a square lattice
in a finite magnetic field $h\ne0$.
For the loop-algorithm we expect to find a rapidly increasing autocorrelation
time and decreasing performance if the product of magnetic field $h$ and 
inverse temperature is much larger than 1. This is due to the fact that
the external field is incorporated into the loop-algorithm via a-posteriori 
acceptance probabilities for each constructed loop.
for $\beta h \ll 1$ these probabilities 
are still large, whereas at $\beta h \approx 1$
they begin to decrease considerably.
%============================================================
\begin{table}
\caption{\sl
1D chain antiferromagnetic Heisenberg model in a magnetic field $h$:
calculation of magnetization $M$ from QMC simulations with 1000000
update-measurement cycles (left: SSE, right: loop-algorithm) for a system with
$\beta=N_s=16$.
$C\cdot\tau$ is the number of elementary update operations per cycle needed
to achieve a mean autocorrelation time $\tau=1$ for the measurements 
of $M$.}
\label{tab_scal1Dh}
\begin{center}
\begin{tabular}{r@{.}lr@{.}lclrl}
\multicolumn{2}{c}{} & \multicolumn{2}{c}{} &\multicolumn{2}{c}{SSE}
& \multicolumn{2}{c}{Loop} \\ 
\multicolumn{2}{c}{$h/J$} & \multicolumn{2}{c}{$\beta\!\cdot\! h$}
&\multicolumn{1}{c}{\rule[-3.5mm]{0mm}{8.5mm}
$\displaystyle \frac{C\!\cdot\!\tau}{C_0\!\cdot\!\tau_0}$} &
\multicolumn{1}{c}{$M$} 
& $\displaystyle \frac{C\!\cdot\!\tau}{C_0\!\cdot\!\tau_0}$ & 
\multicolumn{1}{c}{$M$} \\[1.7ex] \hline\\[-2.2ex]
0 & 02 & 0 & 32 & 1.00 & $0.008(2\pm 5)$ &   1.00 & $0.0083(7\pm 7)$\\
0 & 04 & 0 & 64 & 1.02 & $0.016(4\pm 6)$ &   0.97 & $0.017(6\pm 2)$ \\
0 & 1  & 1 & 6  & 1.22 & $0.057(8\pm 8)$ &   1.84 & $0.057(7\pm 5)$ \\
0 & 2  & 3 & 2  & 1.98 & $0.24(4\pm 2) $ &   7.55 & $0.24(3\pm 2)$ \\
0 & 4  & 6 & 4  & 1.37 & $0.893(8\pm 9)$ & 167.40 & $0.89(4\pm 3)$ \\
1 & 0  & 16& 0  &  0.86 & $2.069(0\pm 8)$ &2338.74 & $2.(24\pm 12)$ \\
\end{tabular}
\end{center}
\end{table}
%============================================================

Indeed, the numerical results in Table \ref{tab_scal1Dh} demonstrate that 
at $\beta h\approx 10$ the loop algorithm cannot be used any more
because the autocorrelation times get too long. For SSE, on the contrary, 
we do not 
expect any negative effect by introducing a magnetic field whose strength
is of the order of the other elementary interactions, $h\approx J$, since no
a-posteriori acceptance decision is neccessary. We rather presume that
performance is slightly worse for $h/J\ll 1$ because 
there are elementary interaction
vertices with very different energy scales. Both predictions are verified
by the data in Table  
\ref{tab_scal1Dh}. For weak fields $h \ll J$ it might be preferable to
construct loops in zero field, and to introduce the field via an
a-posteriori Metropolis decision on whether to accept loops
which change the magnetization, as it is done in the loop algorithm.

For sake of completeness we also show the corresponding data for the 
2-dimensional Heisenberg model in Table  
\ref{tab_scal2Dh}. The results
demonstrate that the different behavior of SSE and loop-algorithm 
described in the one dimensional case is even more severe in two dimensions: 
the loop-algorithm loses its efficiency already at $\beta h\approx 1.5$.
%============================================================
\begin{table}
\caption{
Square lattice antiferromagnetic Heisenberg model in a magnetic field $h$:
calculation of magnetization $M$ from QMC simulations with 1000000
update-measurement cycles for a system with inverse temperature
$\beta J=16$ and $N_s=16^2$ lattice sites. 
$C\cdot\tau$ is the number of elementary update operations per cycle needed
to achieve a mean autocorrelation time $\tau=1$ for the measurements 
of $M$.}
\label{tab_scal2Dh}
\begin{center}
\begin{tabular}{r@{.}lr@{.}lcr@{.}lrl}
\multicolumn{2}{c}{} & \multicolumn{2}{c}{} &\multicolumn{3}{c}{SSE}
& \multicolumn{2}{c}{Loop} \\ 
\multicolumn{2}{c}{$h/J$} & \multicolumn{2}{c}{$\beta\!\cdot\! h$}
&\multicolumn{1}{c}{\rule[-3.5mm]{0mm}{8.5mm}
$\displaystyle \frac{C\!\cdot\!\tau}{C_0\!\cdot\!\tau_0}$} &
\multicolumn{2}{c}{$M$} 
& $\displaystyle \frac{C\!\cdot\!\tau}{C_0\!\cdot\!\tau_0}$ & 
\multicolumn{1}{c}{$M$} \\[1.7ex] \hline\\[-2.2ex]
0 & 02& 0 & 32 & 1.00 & 0 & $22(4 \pm 8)$ &   1.00 & $0.231(3\pm 3)$\\
0 & 04& 0 & 64 & 0.83 & 0 & $47(9 \pm 8)$ &   1.38 & $0.477(7\pm 7)$ \\
0 & 1 & 1 & 6  & 0.94 & 1 & $41(0 \pm 9)$ &   5.61 & $1.42(2\pm 2)$ \\
0 & 2 & 3 & 2  & 0.44 & 3 & $47(0 \pm 7)$ &  34.25 & $3.48(6\pm 7)$ \\
0 & 4 & 6 & 4  & 0.18 & 7 & $73(7 \pm 4)$ &1691.66 & $7.7(8\pm 7)$ \\
1 & 0 & 16& 0  & 0.12 & 22& $10(6 \pm 4)$ &  $---$ & $-----$ \\
\end{tabular}
\end{center}
\end{table}
%============================================================

\begin{figure}
\begin{center}
\vspace{-1mm}
\epsfig{file=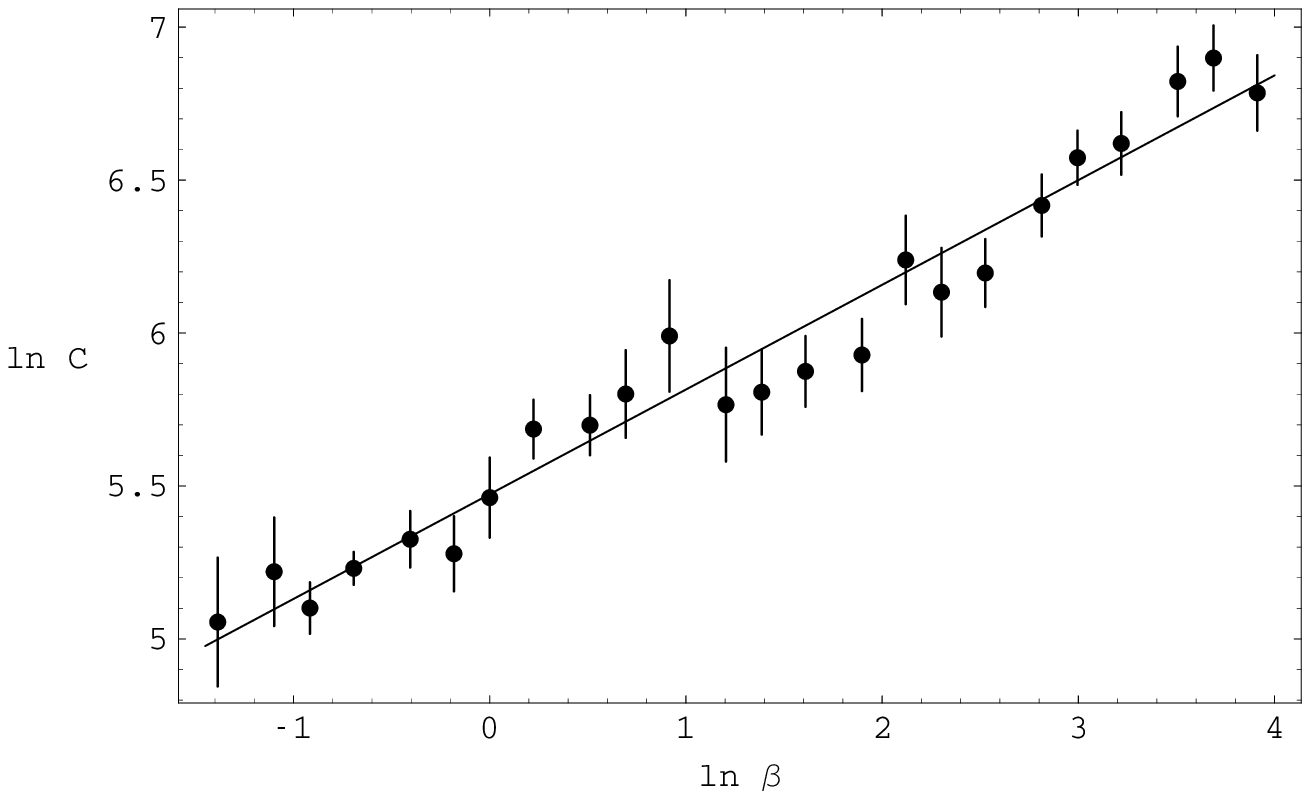,width=7.8cm}
\vspace{-1mm}

\epsfig{file=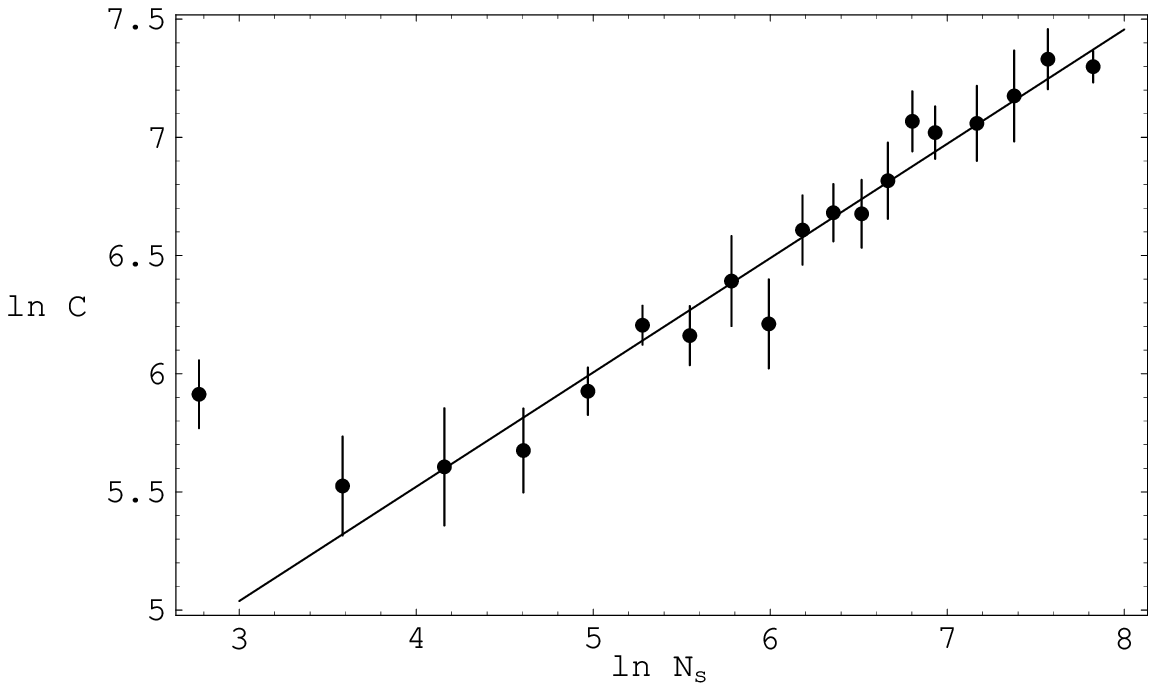,width=7.75cm}
\end{center}
\caption{
Scaling behavior of computation time $C$ to reach a relative accuracy 
of $10^{-4}$ in the measured energy of the 2D AF Heisenberg model.
On top: $\mbox{ln}(C)$ versus $\mbox{ln}(\beta J)$ for $10\times 10$ sites.
Below: $\mbox{ln}(C)$ versus $\mbox{ln}(N_s)$ for $\beta J=10$.
The time was measured in seconds on a DEC workstation.}
 \label{fig_scal_beta_vol}
\end{figure}

In some cases other performance measurements are more interesting.
One could ask how the computation time till a certain
accuracy in a certain measured variable is reached scales with $\beta$ and
$N_s$.
This is studied in Fig.~\ref{fig_scal_beta_vol}.
For the square lattice Heisenberg antiferromagnet 
we trace the computation time to 
reach an accuracy of 4 digits in energy as a function of $\beta$ 
(Fig.~\ref{fig_scal_beta_vol} top) and $N_s$ (Fig.~\ref{fig_scal_beta_vol} 
bottom).
The exponents $\kappa{\scriptstyle (\beta)}$ in 
$C\propto\beta^{\kappa{\scriptstyle (\beta)}}$ and 
$\kappa{\scriptstyle (N_s)}$ in 
$C\propto\beta^{\kappa{\scriptstyle (N_s)}}$ derived from 
Fig.~\ref{fig_scal_beta_vol} are
\begin{eqnarray*}
  \kappa{\scriptstyle (\beta)} &=& 0.34 \pm 0.05, \\
  \kappa{\scriptstyle (N_s)}   &=& 0.48 \pm 0.05. 
\end{eqnarray*}
Both quantities are smaller than 1, 
and Eq.\ (\ref{eq_dyn_exp})
would return a negative dynamical exponents $z$.
This is due to self-averaging:
in a large system local fluctuations of a physical observable around its mean 
value on different subregions of the lattice can compensate and average out
each other, thereby lowering the observable's measured variance.
The computational effort needed to get thermodynamical averages
to a certain relative error scales sublinearly with system size and inverse
temperature, so that systems of several thousand sites or at temperatures
of not more than $0.001 J$ can be simulated within minutes or a few hours
on a standard PC or workstation. 
%=============================================================
%============================================================

%%%%%%%%%%%%%%%%%%%%%%%%%%%%%%%%%%%%%%%%%%%%%%%%%%%%
\section{Green's functions}
\label{sec_GF}
%%%%%%%%%%%%%%%%%%%%%%%%%%%%%%%%%%%%%%%%%%%%%%%%%%%%

The observables listed listed in section \ref{sect_meas}
serve to access important static thermodynamic properties 
of the studied system. However, properties such as photo emission 
$\langle a^\dagger{\sS(\vk,\omega)}\, a{\sS(0,0)} \rangle$ or spin flip
$\langle S^-{\sS(\vk,\omega)}\, S^+{\sS(0,0)} \rangle$ are often even more 
interesting as they provide insights into the system's dynamics.
Within the framework of SSE measuring these Green's functions 
$G(\vk,\omega)$ requires the insertion of local changes on certain 
world-lines (such as removing a particle at propagation level $l_1$ on 
world-line $w_1$ and re-inserting it at propagation level $l_2$ on world-line
$w_2$). Performing these insertions is a highly non-trivial task
since on the one hand detailed balance must be assured, on the other hand the
whole process has to sample all distances $r=w_2-w_1$ and all
propagation differences $\Delta l=l_2-l_1$ efficiently. 
Both requirements are already fulfilled by the loop update steps.
Since this update inserts and moves local changes on the network of world-lines
and connecting interaction vertices it can be used to record the corresponding
Green's functions $G(r,\Delta l)$ ``on the fly'' 
while constructing the loop update.
%=============================================================
\begin{figure}
\begin{center}
\epsfig{file=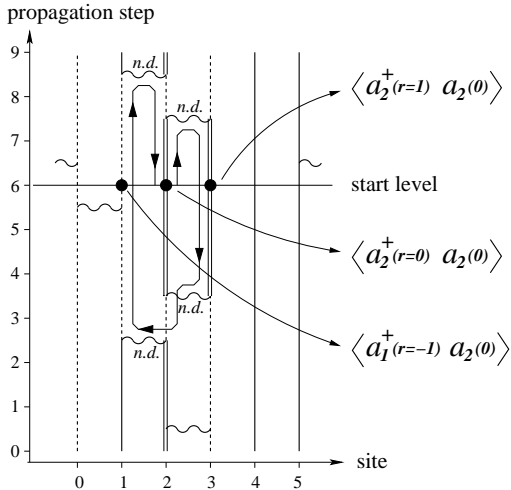,width=6.7cm}
\end{center}
\caption{The loop update constructed in Figs. \ref{fig3} and \ref{fig4} 
can be used to record measurements of the Green's function 
$\langle a^\dagger_2{\sS (r,\Delta l)}\, a^{}_2{\sS (0,0)} \rangle$ and 
$\langle a^\dagger_1{\sS (r,\Delta l)}\, a^{}_2{\sS (0,0)} \rangle$ 
where $r$ is a 
distance between world-lines (or sites) and $\Delta l$ is a propagation level
difference. For sake of clarity only the measurements for $\Delta l=0$ are 
explicitly marked in the figure.}
\label{fig6}
\end{figure}
%============================================================
As an example we reconsider the hard-core boson model from section
\ref{loop_upd} and in particular the operator loop shown in 
Figs. \ref{fig3} and \ref{fig4} which starts with the removal of a 
type-2 particle on propagation level 6 of world-line 2; 
our cut-off power in the series expansion was $L=9$, and 
previous diagonal updates have produced $n=7$ ``non-identity'' interaction 
vertices. 

Taking level 6, the starting point of the loop, as zero point for the 
propagation direction we are now able to measure quantities of type
$\langle a^\dagger_1{\sS (r,\Delta l)}\, a^{}_2{\sS (0,0)} \rangle$ and
$\langle a^\dagger_2{\sS (r,\Delta l)}\, a^{}_2{\sS (0,0)} \rangle$
during the construction of this loop. 
Fig.~\ref{fig6} shows that for $\Delta l=0$  
exactly 2 measurements of $\langle a^\dagger_2{\sS (r,\Delta l=0)} \,
a^{}_2{\sS (0,0)} \rangle$ and one of 
$\langle a^\dagger_1{\sS (r,\Delta l=0)} \, a^{}_2{\sS (0,0)} \rangle$ 
can be performed during the loop: one at the start
(or end) of the loop at distance $r=0$, two on adjacent world-lines ($r=1$)
while moving down (right) and up (left). 
The recorded value at each measurement is the product of the matrix elements
of the creation/annihilation operators inserted at the open ends of the loop under construction. 
We denote the state at propagation level 6 in our example 
before inserting the two creation/annihilation operators as 
$| \alpha(6)\rangle$
and the state after insertion of the operators as $| \tilde{\alpha}(6)\rangle$.
Then the $\Delta l\!=\! 0$ matrix element
$\langle a^\dagger_2{\sS (r\!=\!1,\Delta l\!=\!0)} \,a^{}_2{\sS (0,0)}\rangle$
-- measured when the loop head moves down on world line 3 -- is
\begin{eqnarray*}  
 \langle a^\dagger_2{\sS (r=1,\Delta l=0)} \,a^{}_2{\sS (0,0)} \rangle
 = \langle\;\!\tilde{\alpha}(6) |\;\! a^\dagger_2{\sS (r=1,\Delta l=0)} 
   \,a^{}_2{\sS (0,0)}\;\! | \;\!\alpha(6) \rangle\,.  
\end{eqnarray*}
Stepping down by one more propagation level on world line 3 we can record
the $\Delta l\!\ne\! 0$ matrix element 
\begin{eqnarray*} 
 \langle a^\dagger_2&&{\sS(r=1,\Delta l=-1)}\,a^{}_2{\sS(0,0)}\rangle\\
 &&= \langle\;\!\tilde{\alpha}(5) |\;\! a^\dagger_2{\sS (r=1,\Delta l=-1)}\;\!
   |\;\! \alpha(5) \rangle\, \langle\;\! \tilde{\alpha}(6) | \,a^{}_2{\sS (0,0)} 
   |\;\! \alpha(6) \rangle\,.  
\end{eqnarray*}
Leaving our hardcore boson example behind and returning to the general case 
we conclude this paragraph with the remark
that for the creation or annihilation of a fermion the recorded matrix elements
are always equal to 1, while they can adopt other values for spin flips or
the creation/annihilation of bosons.

Having measured and recorded the quantities $G(r,\Delta l)$ (or a
correlation function $C(r,\Delta l)=\langle \opD_2(r,\tau)\opD_1(0,0) \rangle$)
we still have to perform a couple
of non-trivial transformation steps till we obtain the desired quantities
$G(k,\omega)$ and $C(k,\omega)$ which describe the dynamical 
response of the system to external perturbations.
First we have to relate propagation levels $\Delta l$ to imaginary 
times $\tau$, then a Fourier transform brings us from $r$-space to $k$-space;
finally we need an inverse Laplace transform to step from imaginary time $\tau$
to exitation energy $\omega$. 

%%%%%%%%%%%%%%%%%%%%%%%%%%%%%%%%%%%%%%%%%%%%%%%%%%%%
\section{Efficiently Accessing the system's dynamics}
%%%%%%%%%%%%%%%%%%%%%%%%%%%%%%%%%%%%%%%%%%%%%%%%%%%%

In this section we will discuss efficient implementation strategies for
recording $G(r,\Delta l)$ and for the adjacent transformation steps mentioned
above.

The transformation from propagation levels $\Delta l$ to imaginary time $\tau$
requires the same weight factors as discussed earlier for diagonal correlation
functions:
\begin{eqnarray}
\label{GF_tf}
  G(r,\tau) &=& \sum_{\Delta l=0}^n 
  {L \choose \Delta l}\!\left(\frac{\tau}{\beta}\right)^{\Delta l}\!\! 
  \left(1\!-\!\frac{\tau}{\beta}\right)^{L-\Delta l'}\!\!G(r,\Delta l) \\
  &\equiv&  \sum_{\Delta l=0}^L w(\tau,\Delta l)\; G(r,\Delta l)\,. \nonumber
\end{eqnarray}
where 
\begin{equation}
w(\tau,\Delta l) = 
  {L \choose \Delta l}\!\left(\frac{\tau}{\beta}\right)^{\Delta l}\!\! 
  \left(1\!-\!\frac{\tau}{\beta}\right)^{L-\Delta l}.
\end{equation}

Working in a fixed string size representetion with fixed $L$ 
instead of varying $n$
is more convenient because the binomial weight prefactors 
are fixed during the entire simulation and can easily be calculated once
at the beginning of the simulation.  
 
There are several possible ways to implement the recording of $G(r,\Delta l)$
measurements and the adjacent transformation to $G(r,\tau)$. 
The easiest and at first glance fastest way simply writes all
recorded $G(r,\Delta l)$ data into a two-dimensional array with dimensions
$N_s$ and $L\propto N_s\beta$. The transformation to $G(r,\tau)$ can then 
be performed once at the end of the simulation. However, this method has two
problems. A separate measurement has to be recorded 
each time the loop head
steps up or down by one level on a world-line and whenever it traverses an 
interaction vertex. Recording all these
measurements drastically slows down the loop update process. Second,
 for large systems ($N_s \approx 5000$) and low
temperatures ($\beta\approx 40$) the two-dimensional array needed to 
store $G(r,\Delta l)$ contains about 1 billion
elements and needs more memory than available 
on many computer systems. 

In order to overcome these problems one can replace the ``brute force''
recording of data on {\em all} traversed $(r,\Delta l)$ points by a 
Monte Carlo sampling: in each loop-update a distance $\Delta l$ is 
chosen randomly, according to the probabilities in Eq.\ (\ref{GF_tf}),
for each of the times $\tau$ of interest. Measurements are
then performed only at these $\Delta l$ and transformed directly into $\tau$.

In our code we have adopted a third strategy: we perform {\em all}
possible $G(r,\Delta l)$ measurements (thereby exploiting the fact that
$G(r,\Delta l)$ is constant on the entire world-line fragment between tho
adjacent vertices) and directly transform these into $G(r,\tau)$
at the end of each loop update step.
The transformation after each QMC update step is necessary to keep memory
requirements low.

\begin{figure}
\begin{center}
\epsfig{file=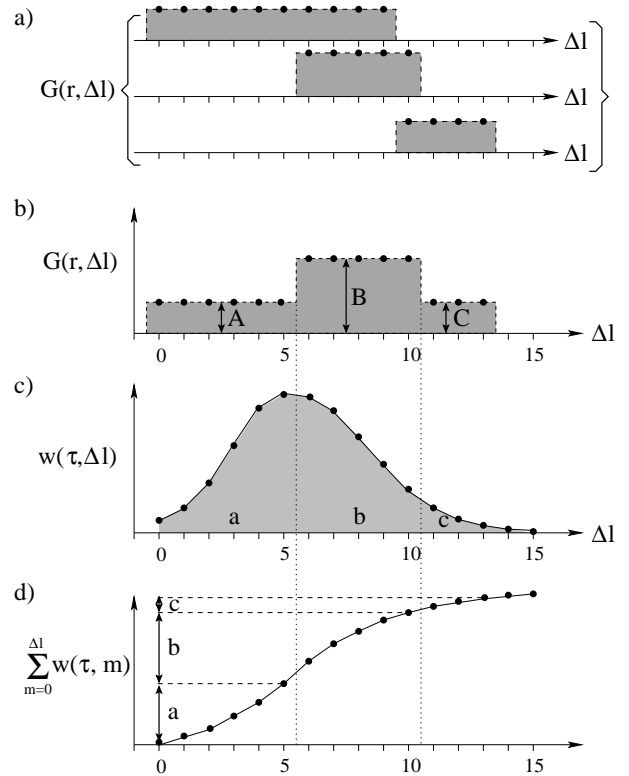,width=8cm}
\end{center}
\caption{Transformation of Green's functions measurements 
from propagation level $\Delta l$ to imaginary time $\tau$: 
the raw measurements
recorded during loop update on different world-line segments (a) are combined 
into a single function $G(r,\Delta l)$ (b). For a given $\tau$ $G(r,\tau)$ 
could be
computed by summing up all $G(r,\Delta l)$ weighted with
$w(\tau,\Delta l) 
= {L \choose \Delta l}\!\left(\frac{\tau}{\beta}\right)^{\Delta l}\!\! 
  \left(1\!-\!\frac{\tau}{\beta}\right)^{L-\Delta l}$ (c).
A much more efficient way uses the ``integrated weight function''
$W(\tau,\Delta l)=\sum_{m=0}^{\Delta l} w(\tau,m)$ (d) to get the total 
contribution 
of each range $]\Delta l_1,\Delta l_2]$ in which $G(r,\Delta l)$ is constant. 
In the example
shown here $G(r,\tau)$ is then simply $aA+bB+cC$.}
\label{fig7}
\end{figure}

Simply applying Eq.\ (\ref{GF_tf}) with its computationally expensive 
operations 
(divisions,powers,binomial coefficients,large sums) would now cost by far 
too much computation time. Instead we remember that $G(r,\Delta l)$ is composed
of a relatively small number of intervals 
$I=]\Delta l_1{\sS (I)},\Delta l_2{\sS (I)}]$ 
with constant function value (Fig.~\ref{fig7}b)). 
Therefore we can compute the contribution of an entire $\Delta l$-interval to
$G(r,\tau)$ in one step:
\begin{equation}
  G(r,\tau) = \sum_{I} G(r,I) \big(W(\tau,\Delta l_2{\sS (I)}) - 
    W(\tau,\Delta l_1{\sS (I)}) \big),
\end{equation}
where $W$ is the ``integrated weight function''
\begin{equation}
  W(\tau,\Delta l) = \sum_{m=0}^{\Delta l} w(\tau,m).
\end{equation}
The $\Delta l$-range in which $W(\tau,\Delta l)$ considerably differs 
from 0 and 1 is determined by mean value and standard deviation of the binomial
distribution $w(\tau,\Delta l)$
\begin{eqnarray}
  \langle \Delta l \rangle &=& L \frac{\tau}{\beta}\\
  \sigma_{\!\Delta l} &=& \sqrt{L \frac{\tau}{\beta}\left(1-\frac{\tau}{\beta}
  \right)}.
\end{eqnarray}
Below $\langle \Delta l \rangle - 5\,\sigma_{\!\Delta l}$ the integrated weight
is zero, above $\langle \Delta l \rangle + 5\,\sigma_{\!\Delta l}$ it is 1
(up to an error of less than $10^{-7}$). The remaining interval rarely
contains more than fifty or hundred $\Delta l$-points (see Fig. 
\ref{fig7}d); these values can easily be stored
after having been computed once for each  
$\tau$. 
Thus $W(\tau,\Delta l)$ can be calculated very rapidly with nothing 
but a couple of ``cheap'' elementary operations.

%=============================================================
%============================================================
For very large systems and very low temperatures
the ``relevant'' $\Delta l$-ranges 
might become so large that it is unfavorable to store all needed
$W(\tau,\Delta l)$ values -- for example because accessing the large array
$W[\tau_i,\Delta l]$ would caust too many cache misses.  
In this case one can store the 
coefficients of some interpolation functions for $W(\tau,\Delta l)$ instead of 
the function values themselves. 
Practical tests have shown that dividing 
the relevant interval $[\langle \Delta l \rangle - 5\sigma_{\Delta l},
\langle \Delta l \rangle + 5\sigma_{\Delta l}]$ into six sub-intervals
with boundaries $\langle \Delta l \rangle - 5\,\sigma_{\!\Delta l}$,
$\langle \Delta l \rangle - 2.8\,\sigma_{\!\Delta l}$,
$\langle \Delta l \rangle - 1.3\,\sigma_{\!\Delta l}$,
$\langle \Delta l \rangle$,
$\langle \Delta l \rangle +1.3\,\sigma_{\!\Delta l}$,
$\langle \Delta l \rangle +2.8\,\sigma_{\!\Delta l}$ and
$\langle \Delta l \rangle +5\,\sigma_{\!\Delta l}$ 
and interpolating $W$ in each sub-interval by a fifth-order polynomial is a 
good compromise between evaluation speed (about 15 elementary operations), 
storage requirements (36 floating point numbers for each  $\tau$) 
and interpolation accuracy (better than $2..3\times 10^{-7}$). 
  
The next transformation step, Fourier transform from $G(r)$ to $G(k)$, 
is a well known standard method that does not impose any fundamental problems. 
However, standard Fast Fourier Transform (FFT) algorithms perform best if
{\em all} $G(k)$ values are to be calculated, whereas in practice one
rarely needs all $k$-values and is interested only in one $k$-point or in some
special points of the Brillouin zone, e.g.~the point $k=(\pi,\pi)$ and its 
immediate neighborhood. Then one can save a lot of computation time by not 
recurring to FFT but using optimized algorithms designed particularly for these
cases. 
If we are interested in only one or a few $k$-points we can use a simple 
Fourier transform to get $\{ G(k,\tau) \}$ from $\{G(r,\tau)\}$ in 
$\cal{O}$$(N_s\cdot n_k)$ operations ($n_k$ is the number of $k$-points).
Correlation functions $C(k,\tau)$ can even be measured directly in $k$-space,
which also can be done in $\cal{O}$$(N_s\!\cdot\! n_k)$ operations.
For the case $1\ll n_k \ll N_s$ we have implemented a new Fourier 
transform algorithm performing
much better than FFT in this situation.\cite{dorn.01}

Unlike a Fourier transform a Laplace transform in general cannot be inverted .
Therefore the last transition step from $\tau$ to $\omega$ is by far more 
complicated than the previous one from $r$ to $k$. 
We use {\it Maximum Entropy} techniques developed within the last years 
and refer to earlier 
publications.\cite{preu.97}

%%%%%%%%%%%%%%%%%%%%%%%%%%%%%%%%%%%%%%%%%%%%%%%%%%%%%%%%%%%
\section{Example: spin correlations of the 2D Heisenberg Model}
%%%%%%%%%%%%%%%%%%%%%%%%%%%%%%%%%%%%%%%%%%%%%%%%%%%%%%%%%%%

In this section we use our standard benchmark model -- the sauqre lattice 
Heisenberg antiferromagnet -- to test our method of Green's functions 
measurements
for correctness and numerical efficiency. To this purpose we calculate
and compare the correlation functions
\begin{eqnarray}
\label{Sz_Sz} 
  &&\langle S^z(\tau,k)\,S^z(0,k)\rangle \quad \mbox{and}\\
\label{S+_S-}
  &&\langle S^+(\tau,k)\,S^-(0,k)\rangle.
\end{eqnarray}
In the $S^z$-eigenbasis, which is normally used to span the model's Hilbert
space, expression (\ref{Sz_Sz}) is a time-dependent correlation function
of two diagonal operators. Therefore the diagonal operator 
in Eq.\ (\ref{Sz_Sz}) can be measured using
Eq.\ (\ref{time_dep_corr}) without introducing changes in the 
world-lines and vertices defining the current state of the system.
The Green's function Eq.\ (\ref{S+_S-}), however, consists 
of two non-diagonal operators and can only
be measured with our new method of recording general Green's functions
that was described in section \ref{sec_GF}.
Furthermore, at zero field $h=0$ both correlation functions are related
via
\begin{equation}
\label{SzSz_S+S-}
  \langle S^z(\tau,k)\,S^z(0,k)\rangle 
  = \frac{1}{2} \langle S^+(\tau,k)\,S^-(0,k)\rangle,
\end{equation}
so that the correctness of both estimators can be checked by directly
comparing these two quantities.

When working with the antiferromagnet wee need to keep in mind that in order 
to keep the exchange matrix elements $S^+S^-$ and $S^-S^+$ positive
we need to perform a gauge transformation, multiplying $S^+$ and $S^-$ on one
sublattice by $-1$. This gauge transformation does not affect any diagonal 
operator, but leads to
a momentum shift of $Q=(\pi,\pi)$, and (\ref{SzSz_S+S-}) 
for the Green's function and Eq.\ (\ref{SzSz_S+S-}) becomes
\begin{equation}
  \label{SzSz_S+S-2}
  \langle S^z(\tau,k)\,S^z(0,k)\rangle 
  = \frac{1}{2} \langle \hat{S}^+(\tau,k+Q)\,\hat{S}^-(0,k+Q)\rangle.
\end{equation}
The numerical data in Table \ref{tab_SzSz_S+S-} perfectly fulfil this 
equality and hence demonstrate the correctness of our Green's functions
measurements. 

%============================================================
\begin{table}
\caption{Comparison of 
$\frac{1}{2} \langle \hat{S}^+(k+Q,\tau)\,\hat{S}^-(k+Q,0)\rangle$ and
$\langle S^z(k,\tau)\,S^z(k,0)\rangle$ for the
$16\times 16$-site 2D AF Heisenberg model at $\beta=16$ and zero magnetic
field. The table shows some $k$-values around $(\pi,\pi)$.}
\label{tab_SzSz_S+S-}
\begin{center}
\begin{tabular}{llrr}\\[-2.3ex]
\multicolumn{1}{c}{$k$} & \multicolumn{1}{c}{$\tau$} & 
\multicolumn{1}{c}{
  $\langle S^z{\scriptstyle (k,\tau)} \, S^z{\scriptstyle (k,0)}\rangle$} &
\multicolumn{1}{c}{ $\!\!\!\!\!\!\!\frac{1}{2}
  \langle S^+\!\!{\scriptstyle (k+Q,\tau)}\,S^-\!\!{\scriptstyle 
(k+Q,0)}\rangle$} \\ 
[0.8ex]\hline\\[-2.3ex]
$(\frac{\pi}{2},\frac{\pi}{2})$  & 0   &$0.16(89\pm 20)$&$0.168(2\pm 4)$ \\
                                 & 0.1 &$0.00(01\pm 17)$&$0.003(2\pm 3)$ \\
                                 & 0.5 &$-0.00(36\pm 15)$&$-0.000(4\pm 3)$ \\ 
[1.2ex]
$(\frac{3\pi}{4},\frac{3\pi}{4})$& 0   & $0.38(39\pm 21)$&$0.386(0\pm 6)$ \\
                                 & 0.1 & $0.01(74\pm 20)$&$0.020(9\pm 5)$ \\
                                 & 0.5 & $0.00(41\pm 19)$&$0.000(4\pm 4)$ \\ 
[1.2ex]
$(\pi,\pi)$                      & 0   & $11.3(35\pm 17)$& $11.36(1\pm 9)$ \\
                                 & 0.1 & $10.4(04\pm 17)$& $10.42(3\pm 9)$ \\
                                 & 0.5 & $9.0(83\pm 17)$& $9.09(3\pm 9)$ \\ 
[1.2ex]
$(\frac{3\pi}{4},\pi)$           & 0   & $0.53(85\pm 23)$& $0.542(2\pm 7)$ \\
                                 & 0.1 & $0.06(31\pm 20)$& $0.063(0\pm 6)$ \\
                                 & 0.5 & $0.00(38\pm 17)$& $0.000(0\pm 5)$ \\ 
[1.2ex]
$(\frac{\pi}{2},\pi)$            & 0   & $0.28(92\pm 23)$& $0.287(6\pm 5)$ \\
                                 & 0.1 & $0.01(08\pm 19)$& $0.008(5\pm 4)$ \\
                                 & 0.5 & $0.00(06\pm 17)$& $0.000(3\pm 4)$ \\ 
\end{tabular}
\end{center}
\end{table}
%============================================================

In the simulation recorded in Table \ref{tab_SzSz_S+S-} we have calculated
$\langle S^z\,S^z\rangle$ and $\langle S^+\,S^-\rangle$ for all allowed 
$k$-points on the path $(0,0)\to(\pi,0)\to(\pi,\pi)\to(0,0)$. Table 
\ref{tab_SzSz_S+S-} shows a subset of these points in the vicinity
of $(\pi,\pi)$.
The three tasks 
``performing updates'', 
``measuring $\langle S^z\,S^z\rangle$'' and 
``measuring $\langle S^+\,S^+\rangle$'' contributed the following 
percentages to overall computation time:
\begin{center}
\setlength{\arraycolsep}{3mm}
\begin{tabular}{lcl}
performing updates &:& 18.8 \% \\
measuring $\langle S^z\,S^z\rangle$ &:& 36.1 \% \\
measuring $\langle S^+\,S^-\rangle$ &:& 45.1 \%
\end{tabular}
\end{center}
From this list and the measurement accuracies in Table \ref{tab_SzSz_S+S-}
we conclude that the highly non-trivial Green's functions measurements lead
to a slightly better accuracy than the direct 
$\langle S^z({\bf r},\tau)S^z({\bf r}',\tau')\rangle$ measurements
while consuming roughly the same amount of computer time as the latter.
Measuring the Green's function is thus the preferred method of
determining also the diagonal real space dynamical correlation functions.

%%%%%%%%%%%%%%%%%%%%%%%%%%%%%%%%%%%%%%%%%%%%%%%%%%%%%%%%
\section{summary}
%%%%%%%%%%%%%%%%%%%%%%%%%%%%%%%%%%%%%%%%%%%%%%%%%%%%%%%%

Stochastic Series Expansion (SSE) together with the implementation tricks and 
Green's functions measurements described in this paper is a highly performant
quantum Monte Carlo simulation technique allowing to access both static and
dynamical properties of very large systems of thousands of sites and at
very low temperatures. Compared to the loop-algorithm, which is slightly
faster on big systems for some specific Hamiltonians, SSE has the advantages
of not suffering from exponential slowing down in external 
fields; furthermore, SSE is more easily applicable to wide classes of 
Hamiltonians.

We thank A.~Sandvik for valuable discussions. This work was supported by DFG
HA 1537/16-1,2 and KONWIHR OOPCV ( A.~D.~) and the Swiss National Science 
Foundation ( M.~T.~). High performance calculations were
performed at HLRZ J\"ulich and LRZ Munich.

%%%%%%%%%%%%%%%%%%%%%%%%%%%%%%%%%%%%%%%%%%%%%%%%%%%%%%%%
 
\end{document}